\documentclass[aps,pre,amssymb,amsmath,twocolumn,floatfix]{revtex4}
\usepackage[dvips]{graphicx}
\begin{document}
\title{Control of accuracy in the Wang--Landau algorithm}
\author{L. Yu. Barash$^{1,2,3}$}
\author{M. A. Fadeeva$^{2,3}$}
\author{L. N. Shchur$^{1,2,3}$}
\affiliation{$^1$ Landau Institute for Theoretical Physics, 142432 Chernogolovka, Russia}
\affiliation{$^2$ Science Center in Chernogolovka, 142432 Chernogolovka, Russia}
\affiliation{$^3$ National Research University Higher School of Economics, 101000 Moscow, Russia}

\begin{abstract}
The Wang--Landau (WL) algorithm has been widely used for simulations in many areas of physics.
Our analysis of the WL algorithm explains its properties and
shows that the difference of the largest eigenvalue of the transition matrix in the energy space
from unity can be used to control the accuracy of estimating the density of states.
Analytic expressions for the matrix elements are given in the case
of the one-dimensional Ising model.
The proposed method is further confirmed by numerical results for the one-dimensional
and two-dimensional Ising models and also the two-dimensional Potts model.
\end{abstract}

\maketitle

\section{Introduction}

The Wang--Landau (WL) algorithm~\cite{Wang-Landau, Wang-Landau-PRE} has been shown
to be a very powerful tool for directly determining the density of states (DOS) and is
also quite widely applicable.
It overcomes some difficulties existing in other Monte Carlo algorithms (such as
critical slowing down) and allows calculating thermodynamic observables,
including free energy, over a wide temperature range in a single simulation.

A number of papers investigated statistical errors of the DOS estimation, and it was found
in~\cite{Yan2003} that errors reach an asymptotic value beyond which additional
calculations fail to improve the accuracy of the results.
Yet it was established in~\cite{Zhou2005,Lee2006}
that the statistical error scales as the square
root of the logarithm of the modification factor,
if the factor is kept constant.

It follows from the results in~\cite{Yan2003} that there is a systematic error
of DOS estimation by the WL algorithm~\footnote{In fact, in the very early presentation
of the algorithm in the Rahman Prize Lecture in 2002, David Landau already mentioned the systematic error in the DOS estimation.}.
It was also confirmed in the case of the two-dimensional Ising model that the deviation of the DOS obtained with the WL algorithm
from the exact DOS does not tend to zero~\cite{1overt,1overt-a}.
Several improvements of the behavior of the modification factor in the algorithm, which were
shown to overcome the problem of systematic error in selected applications, have been suggested~\cite{1overt, 1overt-a,SAMC,SAMC2,Eisenbach2011}.

There are about fifteen hundred papers that apply the WL algorithm and its improvements
to particular problems (e.g., to the statistics of polymers~\cite{Binder09,Ivanov2016}
and to the diluted systems~\cite{Malakis04,Fytas2013}, among many others).

In this paper, we address the question of the accuracy of the DOS estimation.
We report a method for obtaining information on both the convergence of simulations
and the accuracy of the DOS estimation.
We numerically apply our algorithm to the one-dimensional and the two-dimensional
Ising models, where the exact DOS is known~\cite{Beale}, and to the two-dimensional 8-state Potts model, which
undergoes a first-order phase transition.
We also present analytic expressions for the transition matrix in the energy spectrum for the one-dimensional Ising model.

Our approach is based on introducing the transition matrix in the energy space (TMES),
whose elements show the frequency of transitions between energy levels
during the WL random walk in the energy space.
Its elements are influenced by both the random process of choosing a new configurational state
and the WL probability of accepting the new state.

We consider a chain of random updates
(e.g., flips of randomly chosen spins for the Ising model)
of a system configuration.
Each of the updates is accepted with unitary probability.
This random walk in the configurational space is a Markov chain.
Its invariant distribution is uniform, i.e.,
the probabilities of all states of the physical system are equal to each other.
For any pair $\Omega_A$ and $\Omega_B$ of configurations, the probability of an
update from $\Omega_A$ to $\Omega_B$ is equal to the
probability of an update from $\Omega_B$ to $\Omega_A$.
Hence, the detailed balance condition is satisfied.
Therefore,
\begin{equation}
g(E_k)P(E_k,E_m)=g(E_m)P(E_m,E_k),
\label{simplebalance}
\end{equation}
where $g(E)$ is the true DOS and $P(E_k,E_m)$ is
a probability of one step of the random walk to move
from a configuration with the energy $E_k$
to any configuration with the energy $E_m$.
We introduce the notation
\begin{equation}
T(E_k,E_m)=\min\left(1,\frac{g(E_k)}{g(E_m)}\right)P(E_k,E_m),
\label{Texpr}
\end{equation}
which represents nondiagonal elements of the TMES
of the WL random walk on the true DOS.
Relation (\ref{simplebalance}) can be rewritten as
$T(E_k,E_m)=T(E_m,E_k)$.
Therefore, the TMES of the
WL random walk on the true DOS is a symmetric matrix.
Because the matrix is both symmetric and right stochastic,
it is also left stochastic.
This means that the rates of visiting of all energy levels
are equal to each other.

In simulations with a reasonable modification of the WL algorithm,
the systematic error of determining the DOS can be made arbitrarily small.
In this case, we find that the computed TMES approaches a stochastic matrix
as the computed DOS approaches the true value.
There are several interesting conclusions. First, this explains
the criterion of histogram flatness, which is one of the main
features of the original WL algorithm~\cite{Wang-Landau}.
Because the histogram elements are equal to sums of columns in the TMES,
histogram flatness is related to the closeness of the TMES to a stochastic matrix.
Second, it gives a criterion for the proximity of the simulated DOS to the true value.
We introduce the difference of the largest eigenvalue of the calculated TMES
from unity as a parameter. We show that the parameter is closely connected
with the deviation of the DOS from the true value.
We confirm numerically that the deviation of the DOS from the true value 
decays in time in the same manner as our parameter decays.

We are not aware of any other method for determining the accuracy of
a WL simulation without knowing the exact value of the DOS.

The paper is organized as follows.
In Sec.~\ref{AlgSec} we describe the variants of the WL algorithm.
In Sec.~\ref{TMESSec} we introduce the TMES and, in particular,
we describe the behavior of the TMES for the one-dimensional Ising model.
In Sec.~\ref{DiscussionSec} we present our main results and discussion,
including discussion of properties of the TMES, description of the method
and numerical results for the one-dimensional
and two-dimensional Ising models and for the two-dimensional Potts model.

\section{The algorithms}
\label{AlgSec}

Directly estimating the DOS with the WL algorithm
allows calculating the free energy as the logarithm of the partition function
\begin{equation}
Z=\sum_{k=1}^{N_E}g(E_k)e^{-E_k/k_BT},
\label{partition-function}
\end{equation}
where $g(E_k)$ is the number of states (density of states) with the energy $E_k$, $N_E$ is the number of energy levels,
$k_B$ is the Boltzmann constant, and $T$ is the temperature.

The main idea of the WL algorithm is to organize a random walk in
the energy space. We take a configuration of the system with the energy $E_k$,
randomly choose an update to a new configuration with the energy $E_m$,
and accept this configuration with the WL probability
$\min\left(1,\tilde g(E_k)/\tilde g(E_m)\right)$, where $\tilde g(E)$ is
the DOS approximation. The approximation is obtained
recursively by multiplying $\tilde g(E_m)$ by a factor $f$ at each step of the
random walk in the energy space~\footnote{If the new configuration is not accepted,
then the configuration is left unchanged, and the step is counted as the move
to the energy $E_k$, i.e., $\tilde g(E_k)$ is multiplied by the factor $f$.}.
Each time that the auxiliary histogram $H(E)$ becomes sufficiently flat,
the parameter $f$ is modified by taking the square root, $f:=\sqrt{f}$.
Each histogram value $H(E_m)$ contains the number of moves
to the energy level $E_m$. The histogram is filled with zeros
after each modification of the refinement parameter $f$.
It is convenient to work with the logarithms of the values $S(E_k):=\ln\tilde g(E_k)$
and $F:=\ln f$ (to fit the large numbers into double precision
variables) and to replace the multiplication $\tilde g(E_m):=f\cdot\tilde g(E_m)$
with the addition $S(E_m):=S(E_m)+F$.

At the end of the simulation, the algorithm provides only a relative DOS.
Either the total number of states or the number of ground states
can be used to determine the normalized DOS.

It is natural to ask the following three questions:
\begin{enumerate}
\item[Q1]Which condition for the flatness check is optimal?
\item[Q2]How does the histogram flatness influence the convergence of the
DOS estimation?
\item[Q3]Is the choice of the square root rule to modify
the parameter $f$ optimal?
\end{enumerate}

A practical answer to question Q1 was given in the original
algorithm~\cite{Wang-Landau}: keep the flatness within the accuracy
of about 20\%. Choosing an accuracy between 1\% and 20\%
is sometimes useful~\cite{Wust2012}
but can result in a substantial increase of the simulation time~\cite{Wang-Landau-PRE}.
An answer to question Q3 was obtained in two independent
works~\cite{1overt} and~\cite{SAMC}, which introduced 
modifications of the WL algorithm, the
\verb#WL-1/t# algorithm and the stochastic approximation Monte Carlo (\verb#SAMC#) algorithm,
respectively.

There are two phases of the \verb#WL-1/t# algorithm~\cite{1overt}.
The first phase is similar to the WL algorithm except that
every test of the histogram flatness is replaced with a simpler check:
Is $H(E)\ne0$ for all $E$?
The algorithm enters its second phase if $F\le N_E/t$,
where $t$ is the simulation time measured as the number of attempted spin flips.
For $t>t_s$, the histogram is no longer checked and
$F$ is updated as $F=N_E/t$ at each step.
Here $t_s$ is the simulation time when the \verb#WL-1/t# algorithm 
enters the second phase.

Both modified WL algorithms exhibit the same long-range behavior
of the refinement parameter $F$ proportional to $1/t$ for long simulation times~\cite{SAMC,SAMC2}.
This is natural due to the following
conditions of the convergence:
$\sum_{t=1}^\infty F(t)=\infty$ and $\sum_{t=1}^\infty F(t)^\zeta <\infty$ 
for some $\zeta\in (1,2)$~\cite{SAMC,SAMC2}.
The \verb#SAMC# algorithm has an additional parameter $t_0$,
which is the simulation time when the algorithm enters its second phase.
Obtaining the appropriate value of $t_0$ can be quite cumbersome
because the rule of thumb for choosing $t_0$ given in~\cite{SAMC}
is violated even by the $128\times128$ Ising model~\cite{Janke2017}.
The \verb#WL-1/t# algorithm and its further
improvements~\cite{Zhou2008,Swetnam2011,Wust2009} seem to perform
more reliably.
Here, we use the \verb#WL-1/t# algorithm,
although the main obtained results are qualitatively independent of the
modification choice.

\section{Transition matrix in the energy space}
\label{TMESSec}

We calculate the TMES for the WL random walk as follows.
The elements of the TMES $\tilde T(E_k,E_m)$ are probabilities
for the WL random walk to move from a configuration with the energy $E_k$
to a configuration with the energy $E_m$. For simplicity, we consider the
case of the Ising model with periodic boundary conditions
and the energy $E=-\sum_{<i,j>}\sigma_i\sigma_j$,
where the sum ranges pairs of neighboring spins and $\sigma_i=\pm1$.
The number of energy levels accessible for the WL random walk
is $N_E=L/2+1$ for $d=1$ and $N_E=L^2-1$ for $d=2$,
where the even integer $L$ is the linear size of the hypercubic lattice
and $d$ is the lattice dimension.
A WL random move cannot increase or decrease the energy of the configuration
by more than $d$ energy levels, and every column and every row
of the TMES therefore contains no more than $1{+}2d$ nonzero elements.
The nondiagonal elements of $\tilde T(E_k,E_m)$ can be represented as
\begin{equation}
\tilde T(E_k,E_m)=
\min\left(1,\frac{\tilde g(E_k)}{\tilde g(E_m)}\right)P(E_k,E_m),
\label{TTexpr}
\end{equation}
where $k\ne m$.
In general, the structure of the probability $P(E_k,E_m)$
depends on both the system dimension and the local lattice properties
and is rather complicated.

In the case of the one-dimensional Ising chain of $L$ spins with periodic boundary conditions,
the probability to change energy from $E_k$ to $E_m$ in a WL random move is
\begin{equation}
T(E_k,E_m)=\min\left(1,\frac{g(E_k)}{g(E_m)}\right)
\sum_{i=0}^{2k}\frac{N_iQ_i^{E_k\to E_m}}{g(E_k)},
\label{el-mat-1d}
\end{equation}
where $k\ne m$. Here $k$ is the number of couples of domains walls in the configuration,
which determines the energy level $E_k=-\sum_{j=1}^{L}\sigma_j\sigma_{j+1}=-L+4k$,
$N_i(k,L)$ is the number of configurations where
$i$ domains consist of only one spin and $2k{-}i$ domains consist of more than one spin,
and $Q_i^{E_k\to E_m}(L)$ is the probability that a single spin flip
moves the system to the energy $E_m$ from such configurations.
Occupations of the energy levels of the chain are expressed in terms of
binomial coefficients as $g(E_k)=2C_L^{2k}$
because there are exactly $C_L^{2k}$ ways to arrange the $2k$ domain walls.
Therefore, partition function (\ref{partition-function}) is
\begin{equation}
Z_L=2\sum_{k=0}^{L/2} C_L^{2k} e^{(L-4k)/(k_BT)}. \label{Z_L}
\end{equation}

The detailed analytic expressions for $N_i$ and $Q_i$ are presented
in~Appendix~\ref{NiQiSec}. It follows that
\begin{equation}
T(E_k,E_{k+1})=T(E_{k+1},E_k)=\frac{C^{2k}_{L-2}}{\max\left(C^{2k}_L,C^{2k+2}_L\right)}.
\label{tmatrix1d}
\end{equation}

Equation (\ref{tmatrix1d}) can be understood as follows.
The probability of the system to change energy from $E_k$
to $E_{k+1}$ due to a spin flip is equal to the probability
that there are no domain walls adjacent to the spin.
Therefore, $P(E_k,E_{k+1})=C^{2k}_{L-2}/C^{2k}_L$.
Similarly, $P(E_{k+1},E_k)=C^{2k}_{L-2}/C^{2k+2}_L$.
We hence obtain (\ref{tmatrix1d}).

\section{Results and discussion}
\label{DiscussionSec}

\subsection{TMES and the accuracy of the DOS estimation}

The convergence of the \verb#WL-1/t# algorithm follows
from the arguments presented in~\cite{Zhou2008}.
Therefore, there is a final stage of each simulation, where the normalized DOS
remains almost the same and is close to the limiting one.

We note that the condition that $F(t)$ is much smaller than one
in itself does not guarantee that the algorithm is already in its final stage, because
it follows from $\sum_{t=1}^\infty F(t)=\infty$ that
a substantial cumulative change of the DOS due to a long simulation time is possible.
At the same time, a large value of $F(t)$, resulting in a rapid increase
of the calculated DOS, does not guarantee a rapid increase of the normalized DOS.

The normalized DOS remains almost the same during a long simulation time
of the final stage. Therefore, the rate of increase of the logarithm
of the nonnormalized DOS is nearly the same for all energies.
The behavior of the algorithm is close to a Markov chain in the final stage,
and the TMES remains almost the same.
The invariant distribution of the Markov chain
has the property that all energy levels are almost equiprobable, while
different configurations having the same energy may have different
probabilities. Therefore, the TMES is close to a stochastic matrix
in the final simulation stage.
The following proposition also holds: if the TMES
is close to a stochastic matrix, then the obtained normalized DOS is
close to the true DOS (see details in Appendix~\ref{ConvergenceSec}).

The first phase of the \verb#WL-1/t# algorithm aims to obtain
the first crude approximation for the DOS, while the aim of the second phase
(in which the factor $F$ is updated as $F(t)=N_E/t$ at each step)
is to converge to the true DOS.
Both the histogram flatness test in the original WL algorithm
and the test whether all energies have been visited in the \verb#WL-1/t# modification
are quickly passed in the final stage of the calculation because all energies
are almost equally probable.
A much longer simulation time is required to satisfy these tests
in the early calculation stage, when the probabilities of energy levels
differ substantially.

\subsection{The control parameter}

The largest eigenvalue of any stochastic matrix is equal to one, and we
therefore propose to use the difference of the largest eigenvalue of the
TMES from unity computed during the final stage of the WL simulation
as a criterion for the proximity of the DOS to the true value.

We estimate the elements of the TMES in simulations as follows.
The auxiliary matrix $U(E_k,E_m)$ is initially filled with zeros.
The element $U(E_k,E_m)$ is increased by unity after every WL
move from a configuration with the energy $E_k$ to a configuration with the energy $E_m$.
During the simulations, we compute the normalized matrix
$\tilde{T}(E_k,E_m)=U(E_k,E_m)/\tilde{H}$, where
$\tilde{H}=\sum_{k,m}U(E_k,E_m)/N_E$.
The obtained matrix $\tilde{T}$ approaches the stochastic matrix $T$
in the final stage of calculation.
The difference of the largest
eigenvalue $\lambda_1$ of $\tilde{T}$ from unity gives
the control parameter $\delta=\left|1-\lambda_1\right|$.

There are many algorithms for computing the largest eigenvalue of a matrix,
and almost all are suitable for calculating $\delta$.
We used the power method,
also known as power iteration or Von Mises iteration~\cite{poweriteration}.
The algorithm does not compute a matrix decomposition, so
it is quite efficient for large sparse matrices.
It is terminated when a desired accuracy of the eigenvector approximation
is achieved; the eigenvalue estimate is then found by
applying the Rayleigh quotient to the resulting eigenvector.
The method can be used if $\lambda_1$ is
the eigenvalue of largest absolute value and $|\lambda_1/\lambda_2| \ne 1$,
where $\lambda_1,\dots,\lambda_n$ is the list of the matrix eigenvalues
ordered so that $|\lambda_1|\ge|\lambda_2|\ge|\lambda_3|\ge ...\ge|\lambda_n|$.
The absolute value of any eigenvalue of any stochastic matrix 
is less than or equal to unity, therefore, 
the power method is applicable for estimating $\delta$ in the
final stage of the \verb#WL-1/t# algorithm.
It is known that
$|\lambda^{(k)}-\lambda_1|=O(|\lambda_2/\lambda_1|^{2k})$,
where $\lambda^{(k)}$ is the approximation for $\lambda_1$ obtained after $k$
iterations~\cite{Mehl}, so the error asymptotically decreases
by a factor of $|\lambda_1/\lambda_2|^2$ at each iteration.

The TMES is typically a sparse matrix,
and its storage usually requires only $O(N_E)$ of memory.
The matrix-vector multiplications are performed very efficiently
if the matrix is sparse, so each iteration of the power method
requires only $O(N_E)$ operations in this case.
Software libraries such as ViennaCL~\cite{ViennaCL} contain the implementation 
of the power method for sparse matrices.
The power method may require many iterations if $|\lambda_1/\lambda_2|\approx 1$.
However, we note that the eigenvalue needs to be calculated only occasionally.
For example, in our simulations, we calculate $\delta$ only once 
for each integer $n$, where $n\le100\log t<n+1$.
Such a simulation applies the power method only several thousands of times
during a \verb#WL-1/t# calculation with $10^{13}$ spin flips,
so the computing time used for the eigenvalue calculation is negligible.

\subsection{The histogram flatness}
\label{flatnessSec}

We can calculate the normalized histogram
${\cal H}=H(E_m)/\sum_{m}H(E_m)$ as ${\cal H}=\sum_{k}\tilde{T}(E_k,E_m)$.
Hence, the histogram flatness condition is equivalent
to the property that the matrix $\tilde{T}$ is close to stochastic.
Thus, the histogram flatness is closely connected at the final
simulation stage of the \verb#WL-1/t# algorithm with the proximity to the true DOS.

For the original WL algorithm, there is no guarantee that the rate of
increase of the logarithm of the nonnormalized DOS is the same for all energies
in the final stage of the calculation because the parameter modification rule
$F:=F/2$ results in a rapid decay of $F$, and the algorithm hence converges
because the value of $F$ is negligible.
The histogram flatness check is performed with a finite accuracy
such as several percent, which results in a finite accuracy of the calculated DOS.
The choice of high accuracy in the flatness criterion
can result in a slow convergence and a very long simulation time~\cite{Wang-Landau-PRE}.

\subsection{Normalizing the DOS}

\begin{figure*}
\includegraphics[width=0.49\linewidth]{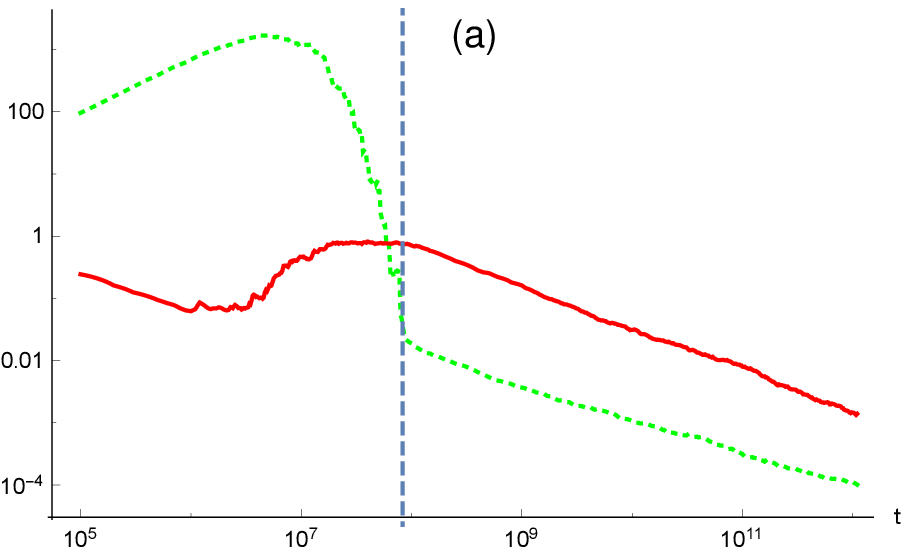}
\includegraphics[width=0.49\linewidth]{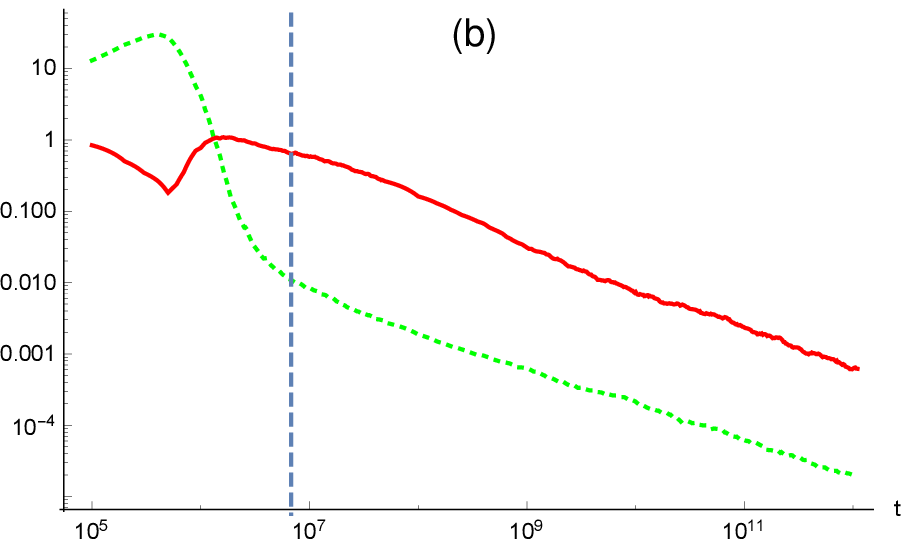}
\caption{Dependence of $\overline{\delta}$ (solid line) and $\overline{\Delta}$ (dotted line)
on the Monte Carlo time $t$ for the \texttt{WL-1/t} algorithm applied to the one-dimensional
Ising model with $L=128$ (left panel) and to the two-dimensional Ising model on the square
lattice of linear size $L=16$ (right panel) and with periodic boundary conditions.
The vertical dashed line marks the average value of $t_s$.
}
\label{Fig12}
\end{figure*}

Normalizing the DOS only at the end of the simulation
was suggested in the original papers~\cite{Wang-Landau,1overt,SAMC}.
We note that this can limit the accuracy of the estimated DOS.
For example, we consider the one-dimensional Ising model with $L=512$,
where the transition to the second phase of the \verb#WL-1/t# algorithm
occurs at $t\sim t_s=2\cdot10^{10}$, where $S(E,t_s)\sim10^7$.
After only several hours of the calculation, we have $t=5\cdot10^{11}$
and $F=N_E/t=5\cdot10^{-10}$. The operation $S(E):=S(E)+F$ is then
beyond the capabilities of double-precision floating-point variables
because there is already a $17$ orders of magnitude difference
between $S(E)$ and $F$. Hence, the operation is
in fact not performed and the DOS is not updated after that.
Therefore, we recommend normalizing the calculated DOS
more frequently during the simulation.
For the simulation corresponding to Fig.~\ref{Fig12},
the calculated DOS is normalized every time
the values of $\delta$ and $\Delta$ are calculated.

\subsection{Behavior of the control parameter for the WL-1/t algorithm}

The parameter
\begin{equation}
\Delta=\frac{1}{N_E}\sum_E
\left|\frac{\tilde{S}(E,t)-S_{\text{exact}}(E)}{S_{\text{exact}}(E)}\right|
\label{Delta}
\end{equation}
estimates the deviation of the computed DOS $\tilde g(E_k)$ 
from the exact DOS $g(E_k)$.
Figure~\ref{Fig12} shows the behavior of $\overline{\Delta}$
and $\overline{\delta}$ as a function of simulation time $t$.
The overline means that the data were obtained by averaging over $M$ independent runs
of the algorithm to reduce statistical noise, where $M=60$ in Fig.~\ref{Fig12}.

We note that $\tilde{S}(E,t)$ in Eq.~(\ref{Delta}) corresponds to the normalized DOS.
Here, we use the normalization $\tilde{S}(E,t)=S(E,t)-\Delta S$,
where $\Delta S=S(E_j,t)-S_{\text{exact}}(E_j)$ and $j$ is chosen
such that $S(E_j)=\max_jS(E_j)$.
Both the abovementioned normalization to the total number of states and
the normalization to the number of ground states turn out
to give values of $\Delta$ close to those presented in Fig.~\ref{Fig12}.
The vertical dashed line marks the average value of $t_s$.

Figure~\ref{Fig12} demonstrates
the monotonic power-law decrease of both the parameters $\delta$ and $\Delta$
during the second phase of the \verb#WL-1/t# algorithm.
We use the logarithmic scale in both axes.
A stable power-law decay of the parameter $\delta$ reveals
the convergence of $\tilde T$ to a stochastic matrix and
can be used as a criterion for the convergence of the simulated DOS to the exact DOS.

\begin{figure*}
\begin{picture}(0,0)
\put(130,12){\includegraphics[width=0.2\linewidth]{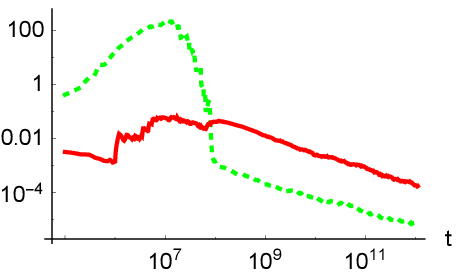}}
\put(380,12){\includegraphics[width=0.2\linewidth]{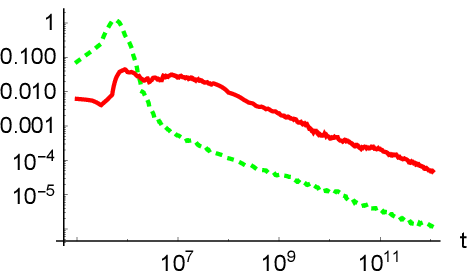}}
\end{picture}
\includegraphics[width=0.49\linewidth]{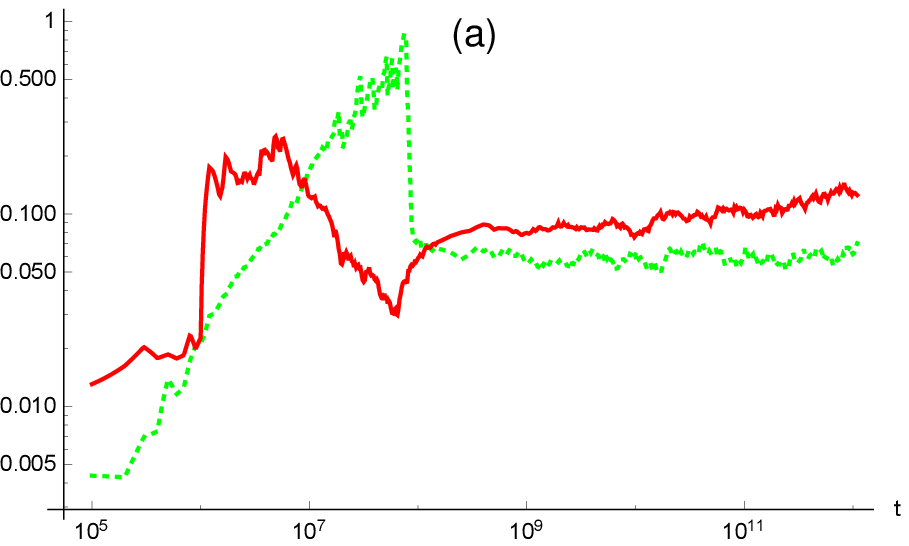}
\includegraphics[width=0.49\linewidth]{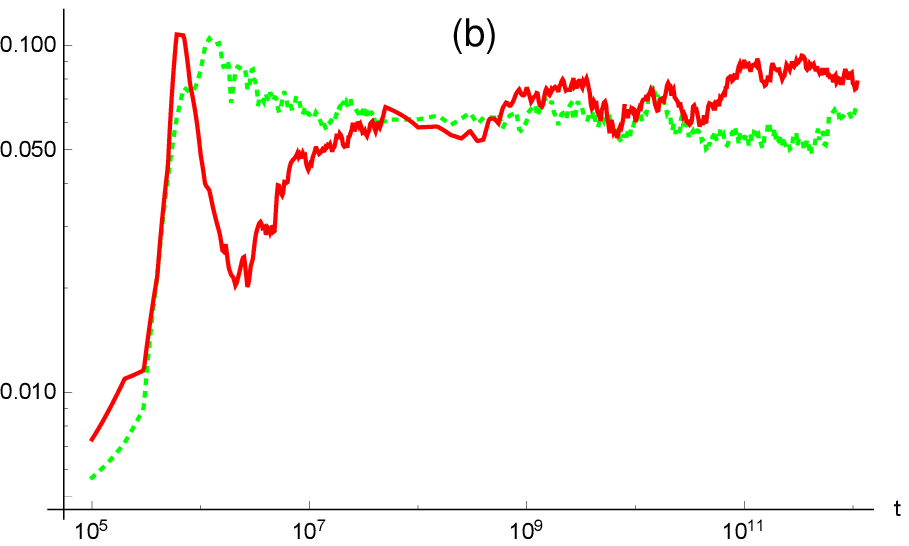}
\caption{ 
Relative standard deviations
$\sigma(\overline{\delta})/\overline{\delta}$ (solid line)
and $\sigma(\overline{\Delta})/\overline{\Delta}$ (dotted line)
as functions of $t$ for the simulations described in Fig.~\ref{Fig12}:
$\sigma(\overline{\delta})$ and $\sigma(\overline{\Delta})$ are standard deviations
of the averaged values $\overline{\delta}$ and $\overline{\Delta}$
obtained using 60 independent runs of the algorithm.
Insets: $\sigma(\overline{\delta})$ (solid line)
and $\sigma(\overline{\Delta})$ (dotted line) as functions of $t$.
}
\label{ErrFig}
\end{figure*}

The fluctuations of the parameters $\delta$ and $\Delta$
are shown in Fig.~\ref{ErrFig} for the simulations described
in Fig.~\ref{Fig12}.
Figure~\ref{ErrFig} shows $\sigma(\overline{\delta})/\overline{\delta}$
and $\sigma(\overline{\Delta})/\overline{\Delta}$ as functions of $t$.
The relative standard deviations were obtained using 60 independent runs of the algorithm.
Therefore, the values in Fig.~\ref{ErrFig} represent the relative magnitudes of the error bars
in Fig.~\ref{Fig12}.
It follows from Fig.~\ref{ErrFig} that $\sigma(\delta)=\sqrt{M}\sigma(\overline{\delta})$ 
and $\sigma(\Delta)=\sqrt{M}\sigma(\overline{\Delta})$ are of the order of
$\overline{\delta}$ and $\overline{\Delta}$, respectively.

\begin{figure}
\includegraphics[width=0.99\linewidth]{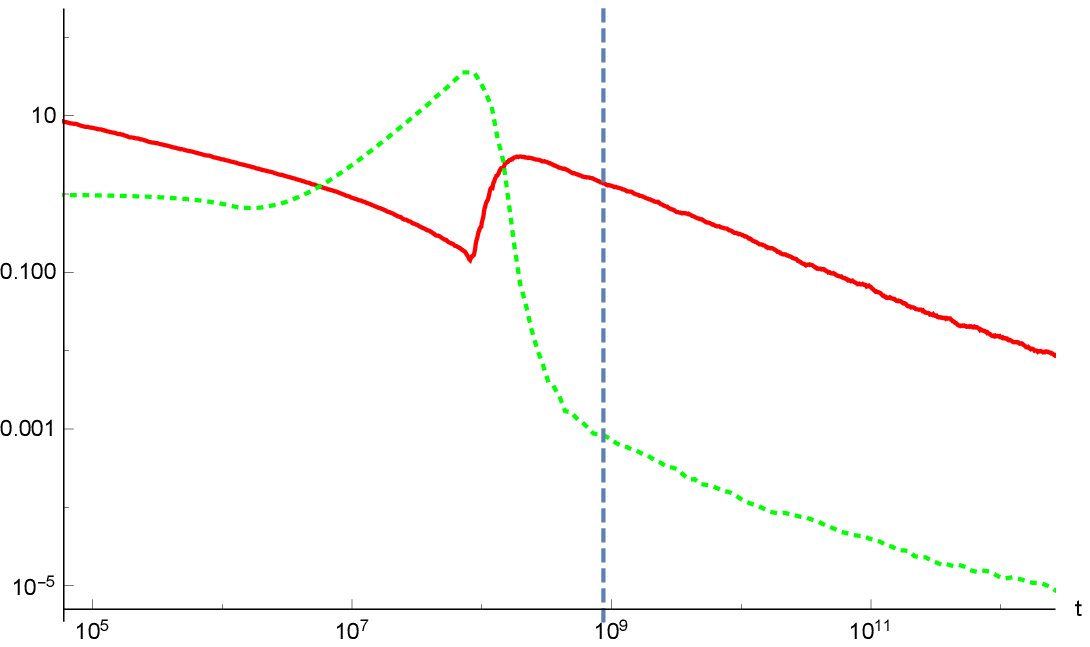}
\caption{
Dependence of $\overline{\delta}$ (solid line) and $\overline{\tilde\Delta}$ 
(dotted line) on the Monte Carlo time $t$
for the \texttt{WL-1/t} algorithm applied to the two-dimensional Potts model
with $q{=}8$ spin states and with periodic boundary conditions.
The lattice size is $L=32$ and $M=40$. Here,
$\tilde\Delta=1/N_E\cdot\sum_E
\left|(\overline{S}(E,t)-S_0(E))/S_0(E)\right|$,
where $S_0(E)=\langle S(E,t=2.6\cdot10^{12})\rangle$.
The vertical dashed line marks the average value of $t_s$.
}
\label{Potts}
\end{figure}

The condition $\delta(t_2)\ll\delta(t_1)$ observed
during the second algorithm phase should result in satisfying
the condition $\Delta(t_2)\ll\Delta(t_1)$,
which allows approximating the value of $\Delta(t_1)$
as the deviation between the DOS computed at $t=t_1$ and $t=t_2$.
This allows estimating the simulation accuracy
in the case where the DOS of the simulated system is not known exactly.
In Fig.~\ref{Potts}, as an example of such a case, we present
the results of simulating the two-dimensional Potts model with $q{=}8$
spin states. The dependence of the parameters $\delta$ and $\tilde\Delta$ on $t$
are qualitatively similar to those calculated for the Ising model (Fig.~\ref{Fig12}).
Because we do not have an analytic expression for the DOS in this case, we calculate
the deviation of $\tilde g(E)$ using the expression $\tilde\Delta=1/N_E\cdot\sum_E
\left|(\tilde{S}(E,t)-S_0(E))/S_0(E)\right|$ and taking $S_0(E)=\tilde{S}(E,t_f)$ for a
large value of $t_f$ ($t_f=2.6\cdot10^{12}$ in Fig.~\ref{Potts}).
The control parameter $\delta$ can thus be used to estimate the accuracy of the obtained DOS.

\begin{figure*}
\includegraphics[width=0.49\linewidth]{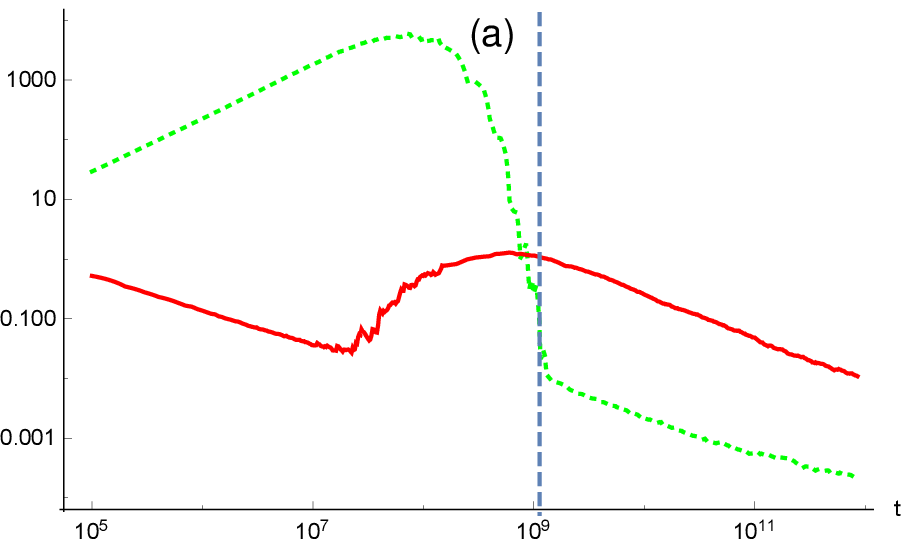}
\includegraphics[width=0.49\linewidth]{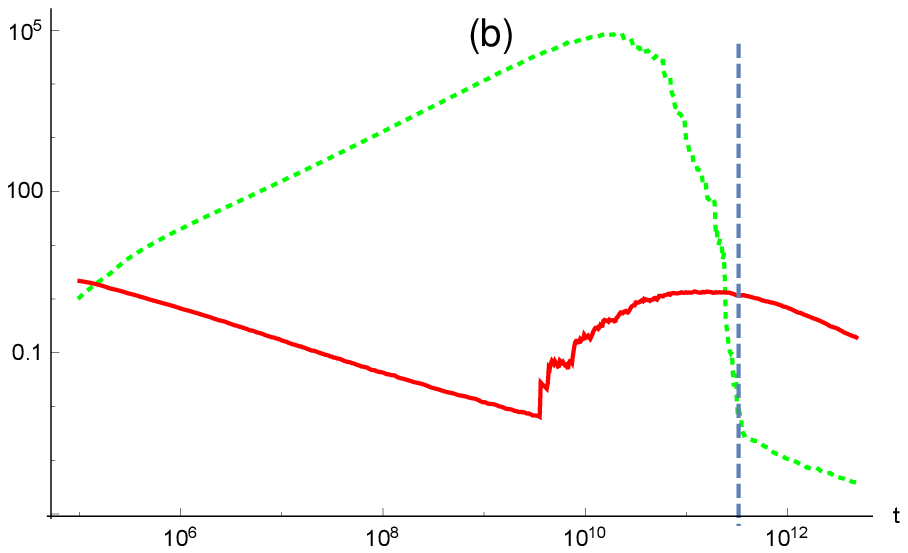}
\caption{
Dependence of $\overline{\delta}$ (solid line) and $\overline{\Delta}$ (dotted line)
on the Monte Carlo time $t$
for the \texttt{WL-1/t} algorithm applied to the one-dimensional Ising model with $L=256$
(left panel) and $L=1024$ (right panel) and with periodic boundary conditions.
The vertical dashed line marks the average value of $t_s$.
}
\label{ising1d}
\end{figure*}

\begin{figure*}
\includegraphics[width=0.49\linewidth]{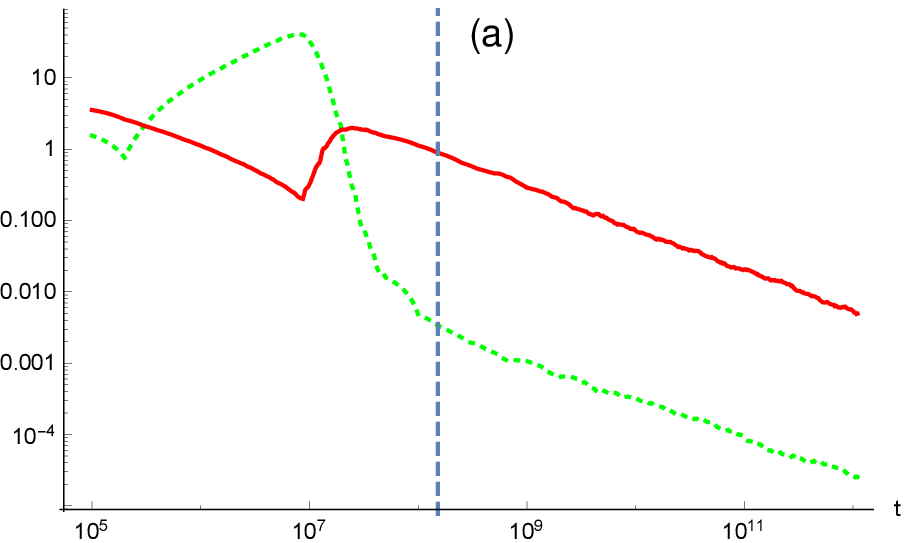}
\includegraphics[width=0.49\linewidth]{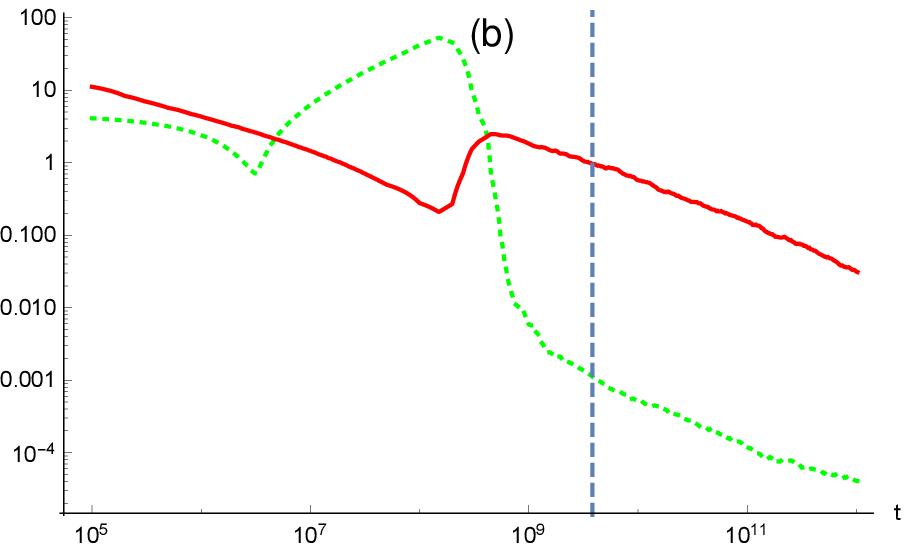}
\caption{
Dependence of $\overline{\delta}$ (solid line) and $\overline{\Delta}$ (dotted line)
on the Monte Carlo time $t$
for the \texttt{WL-1/t} algorithm applied to the two-dimensional Ising model with $L=32$
(left panel) and $L=64$ (right panel) and with periodic boundary conditions.
The vertical dashed line marks the average value of $t_s$.
}
\label{ising2d}
\end{figure*}

Very similar results to those shown in Fig.~\ref{Fig12}
were obtained for various values of the lattice size.
The calculations were performed with $L$ up to 1024 for
the one-dimensional Ising model and up to 64 for the two-dimensional Ising model.
Figures~\ref{ising1d} and~\ref{ising2d} show $\overline{\delta}(t)$ 
and $\overline{\Delta}(t)$ for several different values of the 
Ising model lattice size $L$, where $M=40$.
Figures~\ref{Fig12},~\ref{ising1d} and~\ref{ising2d} 
also demonstrate different values of $t_s$, which grows with the system size.

\subsection{Behavior of the control parameter for the original WL algorithm}

\begin{figure*}
\includegraphics[width=0.48\linewidth]{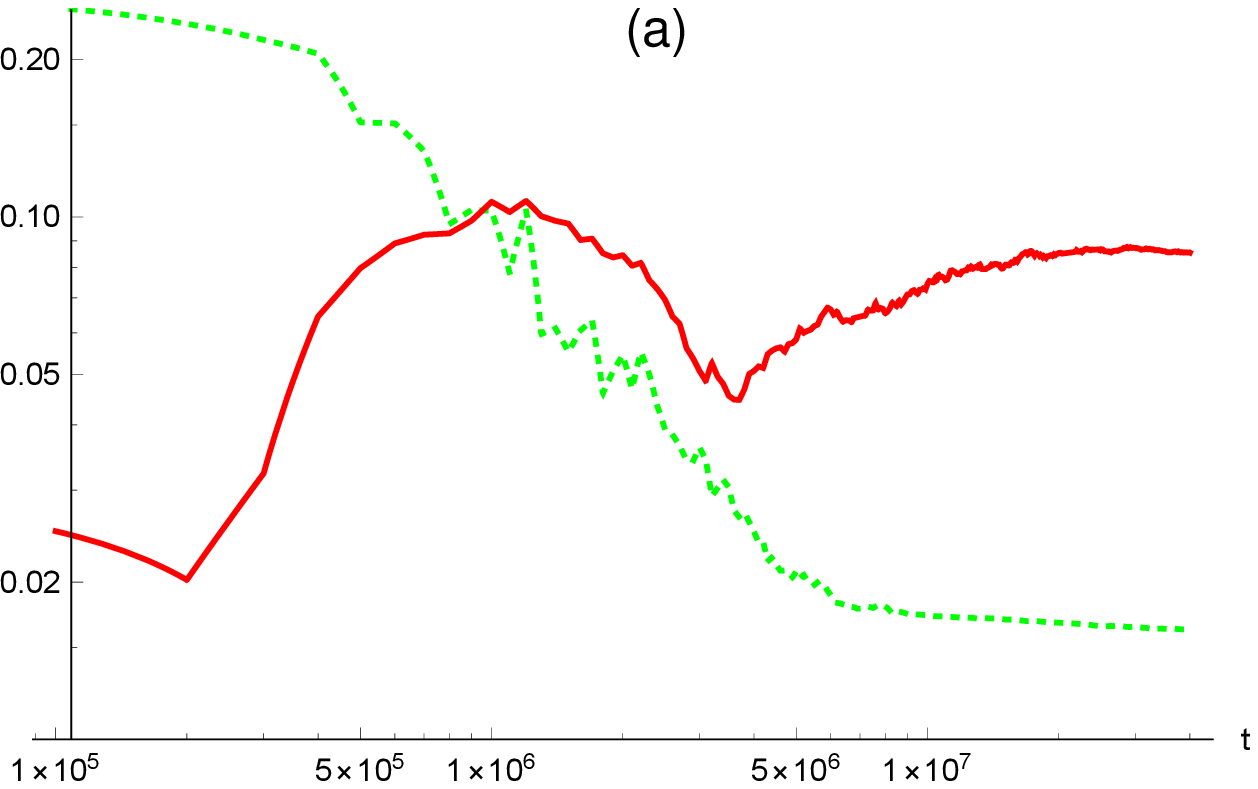}
\includegraphics[width=0.50\linewidth]{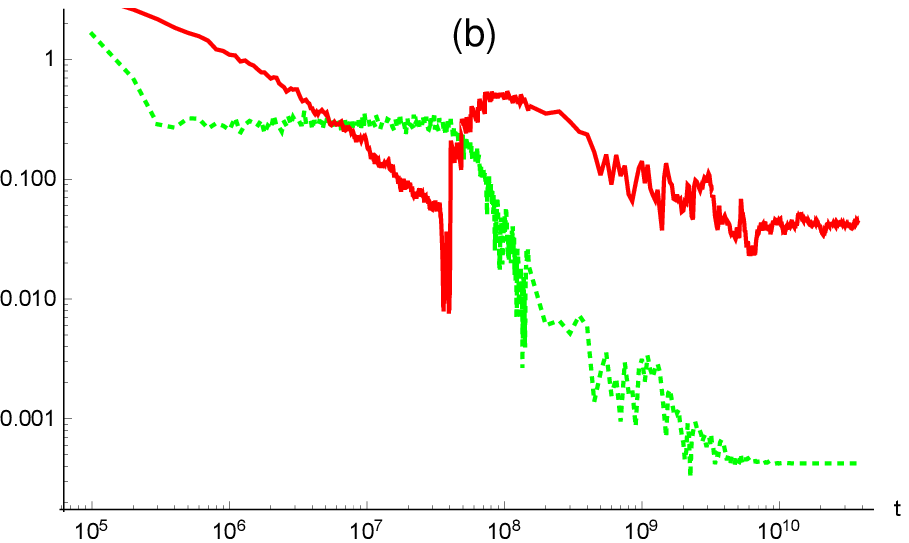}
\caption{
Dependence of $\overline{\delta}$ (solid line) and $\overline{\Delta}$ (dotted line)
on the Monte Carlo time $t$
for the original \texttt{WL} algorithm applied to the one-dimensional Ising model 
with $L=32$ (left panel) and the two-dimensional Ising model with $L=32$ (right panel)
and with periodic boundary conditions.
We use $M=40$ in the left panel and $M=1$ in the right panel.
}
\label{FigClassicalWL}
\end{figure*}

Figure~\ref{FigClassicalWL} shows $\overline{\delta}(t)$ and $\overline{\Delta}(t)$
for the original WL algorithm described in~\cite{Wang-Landau}.
The algorithm was applied to the one-dimensional and two-dimensional Ising models
with $L=32$.
The data in the left panel were obtained by applying the WL algorithm
to the one-dimensional Ising model and averaging over 40 independent runs. 
The right panel
corresponds to a single run of the WL algorithm
applied to the two-dimensional Ising model.

Therefore, both $\Delta$ and $\delta$ saturate for the original
WL algorithm (see also Sec.~\ref{flatnessSec}).
Using the control parameter $\delta$ thus confirms
the systematic error of the original WL algorithm
previously reported 
in~[\citenum{Yan2003},~\citenum{1overt},~\citenum{1overt-a},~24].

\section{Conclusion}
\label{ConclusionSec}

We have analyzed properties of the algorithms and of the TMES.
TMES of the WL random walk on the true DOS is stochastic and symmetric.
We present analytic expressions for the TMES in the case of one-dimensional 
Ising model.
We improve the WL algorithm based on the \verb#WL-1/t# modification
of the original algorithm~\cite{1overt} and propose a method
for examining the convergence of simulations to the true DOS and
for controlling the accuracy of the DOS calculation.
The monotonic power-law decrease of the control parameter $\delta$
during the second phase of the algorithm reveals the convergence of the algorithm,
and the values of the control parameter can be used to estimate the accuracy of
the DOS calculations.

This approach can be generalized to systems with an intitially unknown
discrete spectrum, where the general procedure can be applied for the dynamic change of the TMES.
It would be interesting to check its applicability to systems with a continuous energy spectrum.

\begin{acknowledgments}
This work is supported by the grant 14-21-00158 from the Russian Science Foundation.
\end{acknowledgments}


\appendix
\section{Convergence of the WL-1/t algorithm to the true DOS}
\label{ConvergenceSec}

We have shown that the TMES $T$ of the WL random walk on the true DOS
is stochastic, and also that the TMES $\tilde T$ 
is close to a stochastic matrix in the final stage 
of the \verb#WL-1/t# algorithm.

Here we demonstrate that the obtained normalized DOS is close to the true DOS
if the TMES $\tilde T$ is a stochastic matrix.

It follows from (\ref{TTexpr}) that
\begin{equation}
\frac{\tilde T(E_k,E_m)}{\tilde T(E_m,E_k)}
=\frac{\tilde g(E_k)}{\tilde g(E_m)}
\frac{P(E_k,E_m)}{P(E_m,E_k)},
\end{equation}
where $\tilde g(E)$ is the obtained normalized DOS.
Using (\ref{simplebalance}), we hence obtain
\begin{equation}
\frac{\tilde T(E_k,E_m)}{\tilde T(E_m,E_k)}
=\frac{\eta_m}{\eta_k},
\label{teq}
\end{equation}
where $\eta_i=g(E_i)/\tilde g(E_i)$ and $g(E)$ is the true DOS.
It follows from~(\ref{teq}) and the stochasticity of $\tilde T$ that
\begin{equation}
\eta_m=\eta_m\sum_k\tilde T(E_m,E_k)=\sum_k\tilde T(E_k,E_m)\eta_k.
\label{EqEta}
\end{equation}
Because the TMES is a stochastic matrix,
the rates of visiting all energy levels are equal to each other.
The values of $\tilde g(E)$ therefore remain almost the same,
and the behavior of the algorithm is close to a Markov chain.
Moreover, the invariant distribution of the Markov chain
has the property that all energy levels are equiprobable.
It follows from~(\ref{EqEta}) that
the values $\eta_i/\sum_k\eta_k$
represent the invariant distribution of the Markov chain.
Therefore, $\eta_i$ is independent of $i$, and
the obtained normalized DOS is hence close to the true DOS.

\section{Expressions for $N_i$ and $Q_i$.}
\label{NiQiSec}

We have the relations
\begin{eqnarray}
N_i&=&\frac{L}{k}C_{2k}^iC_{L-2k-1}^{2k-i-1},\;i=0,1,\dots,2k-1,\nonumber\\
N_{2k}&=&2\delta_{L,2k},\label{N_i}\\
Q_i^{E_k\to E_{k-1}}&=&\frac{i}{L},\quad
Q_i^{E_k\to E_k}=\frac{4k-2i}{L},\nonumber\\
Q_i^{E_k\to E_{k+1}}&=&\frac{L-4k+i}{L},\label{Q_i}
\end{eqnarray}
where $\delta_{L,2k}$ is the Kronecker delta.

Expression~(\ref{N_i}) is derived as follows.
We consider the circular chain of $L{-}2k$ spins.
We place the first domain wall in front of the first spin.
We add another $2k{-}i{-}1$ domain walls in the remaining space between the spins;
there are $C_{L-2k-1}^{2k-i-1}$ ways to do this.
Therefore, we have $L{-}2k$ spins and $2k{-}i$ domain walls, where the
first spin of the first domain is the first spin of the chain.

We then add one more spin in every domain.
We also add $i$ domains consisting of only one spin.
There are exactly $C_{2k}^i$ ways to choose $i$ domains among the $2k$ domains.
Each of these choices unambiguously defines
how to add $i$ domains, each consisting of only one spin,
to the available $2k{-}i$ domains of the chain.

We have thus calculated the number of configurations of the circular chain of $L$
spins containing $2k$ domains such that $i$ domains consist of only one
spin, $2k{-}i$ domains consist of more than one spin, and
there is a domain wall in front of the first spin. This number is
$M_i=2C_{2k}^iC_{L-2k-1}^{2k-i-1}$.

When $2k$ domain walls are placed among the $L$ spins, the probability
that there is a domain wall in front of the first spin is equal to $p=2k/L$.
Hence, $N_i=M_i/p$, i.e., we have obtained Eq.~(\ref{N_i}).

The justification of Eqs.~(\ref{Q_i}) is as follows.
We have $2k$ domains, where $i$ domains consist of only one spin
and $2k{-}i$ domains consist of more than one spin.
To remove a couple of domains with just a single spin flip,
we must choose one of $i$ spins from the domains consisting of only one spin.
Therefore, $Q_i^{E_k\to E_{k-1}}=i/L$.

To add a couple of domains with just a single spin flip,
we must choose a spin that is not a boundary spin of a domain.
There are $L{-}4k{+}i$ spins satisfying this condition because
there are $2k$ spins located to the right of a domain wall,
$2k$ spins located to the left of a domain wall,
and $i$ spins which are located with a domain wall on both the right and the left.
Therefore, $Q_i^{E_k\to E_{k+1}}=(L-4k+i)/L$.
Finally, $Q_i^{E_k\to E_k}=1-Q_i^{E_k\to E_{k-1}}-Q_i^{E_k\to E_{k+1}}=(4k-2i)/L$.


\begin{thebibliography}{99}
\frenchspacing
\bibitem{Wang-Landau} F. Wang, D. P. Landau, Phys. Rev. Lett. {\bf 86}, 2050 (2001).
\bibitem{Wang-Landau-PRE} F. Wang, D. P. Landau, Phys. Rev. E {\bf 64}, 056101 (2001).
\bibitem{Yan2003} Q. Yan, J. J. de Pablo, Phys. Rev. Lett. {\bf 90}, 035701 (2003).
\bibitem{Zhou2005} C. Zhou, R. N. Bhatt, Phys. Rev. E {\bf 72}, 025701 (2005).
\bibitem{Lee2006} H.W. Lee, Y. Okabe, and D.P. Landau, Comp. Phys. Comm. {\bf 175} 36 (2006).
\bibitem{1overt} R. E. Belardinelli and V. D. Pereyra, Phys. Rev. E {\bf 75}, 046701 (2007).
\bibitem{1overt-a} R. E. Belardinelli and V. D. Pereyra, J. Chem. Phys {\bf 127}, 184105 (2007).
\bibitem{SAMC} F. Liang, C. Liu, and R. J. Carroll, J. Am. Stat. Ass. {\bf 102}, 305 (2007).
\bibitem{SAMC2} F. Liang, J. Stat. Phys. {\bf 122}, 511 (2006).
\bibitem{Eisenbach2011} G. Brown, Kh. Odbadrakh, D. M. Nicholson, M. Eisenbach, Phys. Rev. E {\bf 84}, 065702(R) (2011).
\bibitem{Binder09} M.P. Taylor, W. Paul, and K. Binder, J. Chem. Phys. {\bf 131}, 114907 (2009).
\bibitem{Ivanov2016} S.V. Zablotskiy, V.A. Ivanov, and W. Paul, Phys. Rev. E {\bf 93}, 063303 (2016).
\bibitem{Malakis04} A. Malakis,A. Peratzakis, and N. G. Fytas, Phys. Rev. E {\bf 70}, 066128 (2004).
\bibitem{Fytas2013} N. G. Fytas and P.E. Theodorakis, Eur. Phys. J. B {\bf 86}, 30 (2013).
\bibitem{Beale} P. D. Beale, Phys. Rev. Lett. {\bf 76}, 78 (1996).
\bibitem{Wust2012} T. W\"ust, D. P. Landau, J. Chem. Phys. {\bf 137}, 064903 (2012).
\bibitem{Janke2017} S. Schneider, M. Mueller, W. Janke, Comp. Phys. Comm. {\bf 216} 1 (2017).
\bibitem{Zhou2008} C. Zhou, J. Su, Phys. Rev. E {\bf 78}, 046705 (2008).
\bibitem{Swetnam2011} A. D. Swetnam, M. P. Allen, J. Comput. Chem. {\bf 32}, 816 (2011).
\bibitem{Wust2009} T. W\"ust, D. P. Landau, Phys. Rev. Lett. {\bf 102}, 178101 (2009).
\bibitem{poweriteration} R. von Mises and H. Pollaczek-Geiringer,
Praktische Verfahren der Gleichungsaufl\"osung,
ZAMM - Zeitschrift f\"ur Angewandte Mathematik und Mechanik {\bf 9}, 152 (1929).
\bibitem{Mehl} 
S. B\"orm, C. Mehl,
Numerical Methods for Eigenvalue Problems, Walter De Gruyter, Berlin/Boston, 2012.
\bibitem{ViennaCL}
K. Rupp, Ph. Tillet, F. Rudolf, J. Weinbub, A. Morhammer, 
SIAM J. Sci. Comp. {\bf 38}, S412 (2016).

\end{thebibliography}
\end{document}